# Influence of Side Chain Isomerism on the Rigidity of Poly(3-alkylthiophenes) in Solutions Revealed by Neutron Scattering

William D. Hong[1], Christopher N. Lam[1], Yangyang Wang[2], Dongsook Chang[2], Youjun He[2], Luis E. Sánchez-Díaz[3], Changwoo Do[1], and Wei-Ren Chen[1, *]

[1]Neutron Scattering Division, Oak Ridge National Laboratory, Oak Ridge, Tennessee 37831, USA.

[2]Center for Nanophase Materials Sciences, Oak Ridge National Laboratory, Oak Ridge, Tennessee 37831, USA.

[3] Department of Chemistry and Physics, The University of Chattanooga, 615 McCallie Ave, Chattanooga, TN 37403

**ABSTRACT:** Using small angle neutron scattering, we conducted a detailed structural study of poly(3-alkylthiophenes) dispersed in deuterated dicholorbenzene. The focus was placed on addressing the influence of spatial arrangement of constituent atoms of side chain on backbone conformation. We demonstrate that by impeding the π- π interactions, the branch point in side chain promotes torsional motion between backbone units and results in greater chain flexibility. Our findings highlight the key role of topological isomerism in determining the molecular rigidity and are relevant to the current debate about the condition necessary for optimizing the electronic properties of conducting polymers via side chain engineering.

Conjugated polymers are synthetic macromolecules that are characterized by a backbone consisting of alternating double- and single-bonds.[1-4] This conjugated structure provides overlapping $p$-orbitals for delocalised $\pi$-electrons and allows them to be electrically conducting upon properly doping.[5] Conjugated polymers are found in a variety of applications.[6-9] The most exploited one is their use as organic electronics including photovoltaic cells and organic field-effect transistors[10-11]; they are also proposed for many other applications, such as in the molecular imaging[12] and pharmaceutical fields[13].

By incorporating soluble side chains onto the backbone, one can increase the solubility of conjugated polymers in common organic solvents to facilitate the processing.[14-15] Among all the developed solution-processable conjugated polymers, poly(3-alkylthiophenes) (P3ATs) represent one of the most studied systems because of its exceptional chemical stability, mechanical strength, optoelectronic properties.[16-17] A direct connection between the chain conformations of P3ATs in their solution states and optoelectronic properties in bulk has been recognized.[9] Unlike traditional flexible polymers which can be described as random coils when dispersed in solutions,[18] conjugated polymers are characterized by more stiff conformations due to the significant excluded volume interactions between the neighboring monomers and strong inter chain interactions.[9, 19] The prospect of optimizing the electronic properties of P3ATs by controlling their conformation has motivated a series of scattering studies to investigate the structural origin of backbone stiffness.[21-28] Multiple factors, including molecular weight and its distributions, side chain chemistry, regioregularity, and backbone defects, have been demonstrated to hinder the chain free rotation and contribute to the chain stiffness of P3ATs.

One intriguing aspect of tuning backbone conformation is through varying side chain isomerism. The difference in the spatial arrangement of constituent atoms of thiophene is expected to change the interactions between neighboring monomers and surroundings. The critical question of how the molecular stiffness of P3ATs depends on the stereoregularity of their chain pendant groups has not been answered. Motivated by this challenge, in this report we used small angle neutron scattering (SANS) to investigate the influence of side chain molecular structure on the backbone conformation of two P3AT systems, the commonly studied poly(3-hexylthiophene) (P3HT) and poly(3-(2-methylpentyl)thiophene) (P3(2MP)T) respectively. As indicated by the insets of Figure 1, they have the same chemical formula but different structural formulas. One common index parameter to quantify the molecular stiffness is the persistence length $b$, which separates the small length scales below which the chain is viewed as a stiff cylinder and the large scales over which a chain takes the shape of a random walk with a unit step of $b$. By analyzing the SANS spectra collected from the solutions of P3HT and P3(2MP)T, we demonstrated that $b$ of P3(2MP)T is clearly higher than that of P3HT under the same conditions.

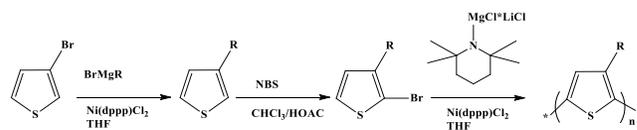

**Scheme 1.** Scheme illustrating the synthesis of P3HT and P3(2MP)T.

P3HT and P3(2MP)T were synthesized at the Center for Nanophase Materials Sciences (CNMS), Oak Ridge National Laboratory (ORNL) by a modified GRIM method.[29-30] The synthetic route was shown in Scheme 1. Briefly, in an inert gas environment, 3-bromothionphene and alkyl magnesium bromide could conduct Kumada reaction[31] with Ni(dppp)Cl$_2$ as the catalyst to yield 3-alkylthiophene compounds. After adding a bromide in the second position by reacting with NBS in a mixture solvent of chloroform and ethyl acetic acid, 2-Bromo-3-alkythiophene monomers were obtained. 2-Bromo-3-alkylthiophenes will react with Grignard Reagent (2, 2, 6, 6-Tetramethylpiperidinylmagnesium chloride lithium chloride complex solution) to give 2-Bromo-5-magnesiumchloridelithiumchloride-3-alkylthiophenes, after using Ni(dppp)Cl$_2$ as catalyst, which will polymerize to give the final polymer products. Both polymers are characterized by a very high regularity which is >97%. From GPC in chloroform, the number average molecular weight Mn of P3HT for P3(2MP)T are found to be 18.4 kg/mol with polymer diversity indexes (PDI) of 1.09 and 33.0 kg/mol with PDI of 1.13. Both polymers were further characterized by NMR and UV-Vis.

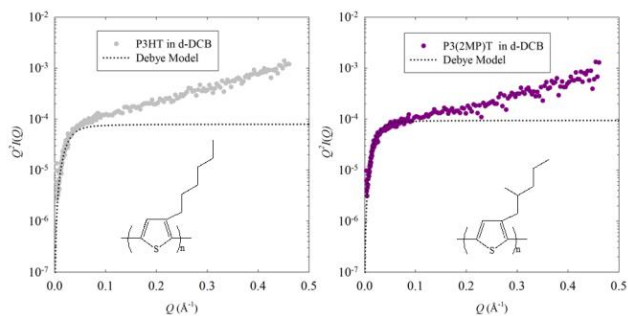

**Figure 1.** The Kratky plots of SANS coherent intensities of P3HT (left panel) and P3(2MP)T (right panel) in d-DCB. The molecular structures of conjugated polymers are given in the insets.

SANS measurements were carried out at Extended $Q$-Range Small-Angle Neutron Scattering Diffractometer (EQSANS) at Spallation Neutron Source (SNS), ORNL. The scattering wavevector $Q$ was ranging from 0.005 to 0.5 $\text{Å}^{-1}$. The samples were contained in Hellma quartz cells with a 2 mm pathlength. The measured data was corrected for detector background, sensitivity, scattering contributed from the empty cells. The intensities were placed on an absolute scale according to a standard procedure provided by the instrument scientists. P3HT and P3(2MP)T were dispersed in o-dicholorbenzene-d4 (d-DCB) with a concentration of 2 $mg/mL$ to minimize the contribution inter-chain spatial correlation to the collected SANS signal. The solutions were kept at 35 °C for overnight under inert atmospheres. We have also tested other concentrations to check the aggregation behavior.

Before presenting the quantitative conformational characteristics obtained from SANS data analysis, it is instructive to investigate the qualitative features of the SANS spectra: Figure 1 gives the Kratky plots of the SANS coherent intensities of P3HT (left panel) and P3(2MP)T (right panel) in deuterated dichlorobenzene (d-DCB). In this representation, the molecular swelling of P3HT and P3(2MP)T conjugated polymers are reveals by the high-$Q$ upturns of the experimental results (colored symbols) deviating from the predicted plateau of Debye model (dotted curves).[32] This observed non-Gaussian conformational feature is due to the delocalization of the $\pi$ electrons migrating along the main chain. In Figure 2 we present the SANS absolute intensities $I(Q)$ obtained from the d-DCB solutions of P3HT (gray symbols) and P3(2MP)T (purple symbols) along with the curves from model fitting (black solid lines). Within the probed temperature range from 25 °C to 77 °C, both sets of $I(Q)$ are seen to be in quantitative agreement with each other within experimental error when $Q > 0.03 \text{ Å}^{-1}$. In both solutions, when $Q > 0.03 \text{ Å}^{-1}$ the incoherent backgrounds $I_{inc}$ essentially show no dependence on temperature. Their magnitudes are found to be around 0.05 $cm^{-1}$. Since $I_{inc}$ is dominated by the number of protons present in the solutions, this observed invariance again provides further evidence that the weight fractions of both samples are indeed identical to each other. In the lower $Q$ region, the $I(Q)$ obtained from both solutions become discernibly different, and the magnitude of $I(Q)$ obtained from the P3(2MP)T solution is seen to be higher than that of the P3HT solution. Given that P3HT and P3(2MP)T have the same scattering length density due to their same chemical compositions and negligible inter-chain interactions at this dilute concentration, this observation indicates the difference in the molecular weight of the conjugated polymers.

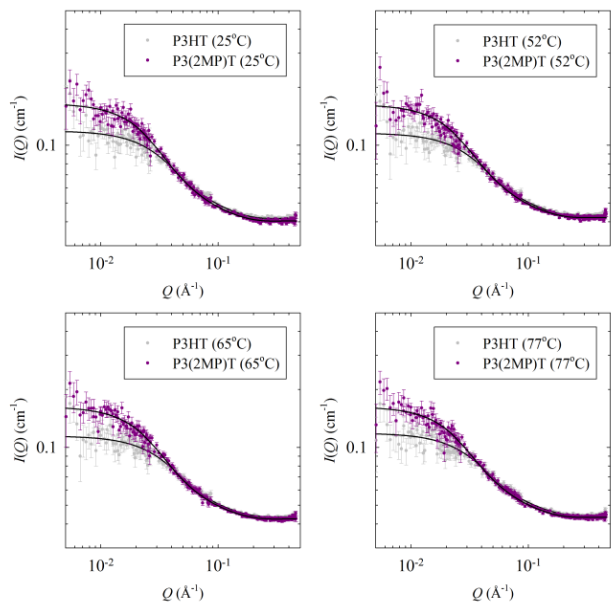

**Figure 2.** Comparison of the SANS absolute intensities $I(Q)$ obtained from the deuterated dichlorobenzene solutions of P3HT (gray symbols) and P3(2MP)T (purple symbols) with a fixed concentration of 2 mg/mL at four different temperatures. Solid curves are the results of model fitting.

Having addressed the generic picture of the polymer conformations from the SANS spectra presented in Figures 1 and 2, one can further explore the structural details by examining the results of model fitting. In this study the SANS absolute intensity $I(Q)$ is modeled by the following expression:[33]

$$I(Q) = \frac{c}{N_A}\left(\frac{\Delta\rho}{d}\right)^2 MP_{SAC}(Q)R(Q) + I_{inc}, \quad (1)$$

where $c$ is the weight fraction of conjugated polymers in the solution, $N_A$ is the Avogadro constant, $\Delta\rho$ is the difference in the bound scattering length density between polymer and solvent, $d$ is the density of polymer, $M$ is the molecular weight of a conjugated polymer chain, and $I_{inc}$ is the incoherent scattering intensity, which is independent of $Q$. $R(Q)$ is the scattering function for the cross section of a cylinder with a radius of $R_{CX}$ and is given by

$$R(Q) = \left[\frac{2J_1(QR_{CX})}{QR_{CX}}\right]^2, \qquad (2)$$

where $J_1(x)$ denotes the Bessel function of the first kind. $P_{SAC}(Q)$ is the form factor of a conjugated polymer chain which describes the intra-chain spatial correlation. As demonstrated in Figure 1, the scattering behaviors of P3HT and P3(2MP)T are not in an agreement with the prediction of the Debye model. To address the influence of excluded volume on a conjugated polymer coil, the scattering function for the form factor of a self-avoiding chain, $P_{SAC}(Q)$,[34] is used in this study. Readers are referred to reference for full numerical details of this model.

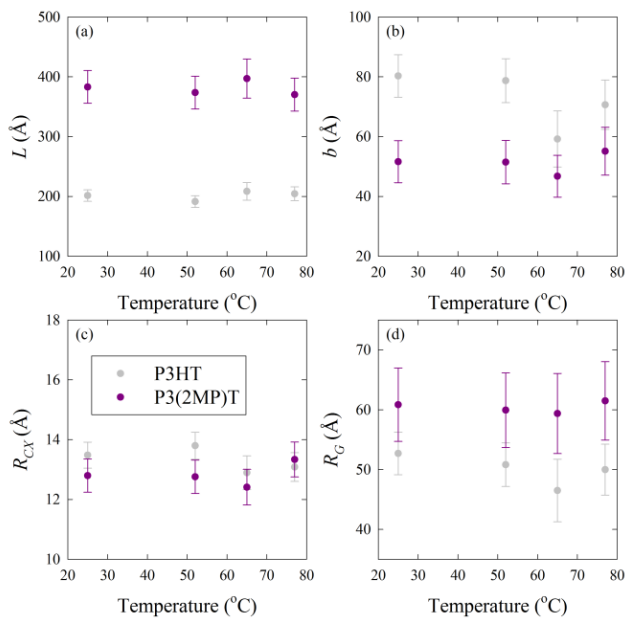

**Figure 3.** Conformational characteristics of P3HT and P3(2MP)T as a function of temparature obtained from SANS model fitting: (a) contour length $L$; (b) persistence length $b$; (c) radius of chain cross section $R_{CX}$; and (d) radius of gyration $R_G$.

The results of model fitting are presented in Figure 3. Figure 3a presents the contour length $L$ as a function of temperature. $L$ is found to be around 200 Å and 370 Å for P3HT and P3(2MP)T, respectively, and both remain essentially unchanged within the probed temperature range. As indicated by Figure 2, the difference in $I(Q)$ observed in the low $Q$ regime is due to the difference in $M$. Since $P_{SAC}(0) = 1$, Eqn. (1) shows that $M$ can be calculated from the extrapolated zero-angle scattering intensity $I(0)$. From the chemical formula of P3HT, P3(2MP)T, and d-DCB, $\Delta\rho$ is calculated to be approximately -3.62 x 10$^{-6}$ Å$^{-2}$ for both P3HT and P3(2MP)T, and $d$ is calculated to be 1.576 and 1.661 g cm$^{-3}$ for P3HT and P3(2MP)T, respectively. Accordingly, SANS data analysis shows that $M$ for P3HT and P3(2MP)T are 20.9 ± 0.7 kg/mol and 38.2 ± 0.7 kg/mol, respectively. These results are seen to be in a reasonable quantitative agreement with those determined by GPC. Figure 3b shows results of the persistence length $b$, which describes the local stiffness of a conjugated polymer chain. P3HT appears to be more rigid than P3(2MP)T when suspended in d-DCB. Moreover, as demonstrated by Figure 3c, the cross section of a P3HT chain is seen to be identical to that of P3(2MP)T. The radius of gyration, $R_G$, has been commonly used to quantify the global size of a polymer chain. Due to the noisy $I(Q)$ in the low $Q$ regime, extracting $R_G$ from the Guinier approach[35] results in significant uncertainty. Alternatively, $P_{SAC}(Q)$ allows one to calculate the $R_G$ of P3HT and P3(2MP)T from the corresponding $L$ and $b$ with satisfactory statistics, and the results are given in Figure 3d. As expected, $R_G$ of P3(2MP)T is seen to be larger than that of P3HT due to the difference in molecular weight. Again, no discernible temperature dependence is observed.

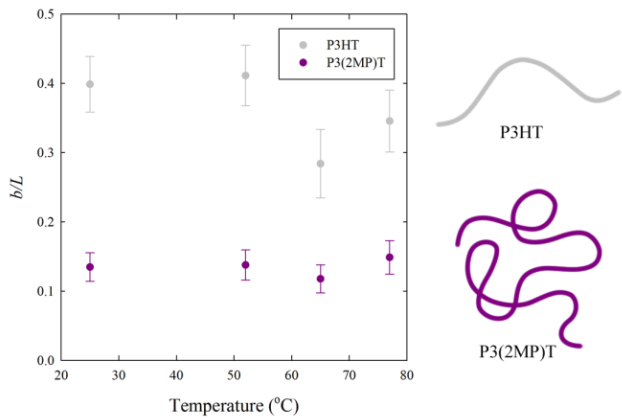

**Figure 4.** The ratio of $b/L$ for P3HT and P3(2MP)T as a function of temperature and the schematic representations of the global conformation of P3HT and P3(2MP)T.

The origin of the observed difference in $b$ presented in Figure 3b is a consequence of the different side chain chemistries of P3HT and P3(2MP)T. Numerous investigations have previously shown that synthetic pathway, regioregularity, and side chain chemistry can affect the persistence lengths of P3ATs. P3ATs have been shown to form crystals with a lamellar structure of stacks of planar thiophene chains that are uniformly spaced by alkyl side chains. Therefore, factors such as synthetic pathway and defects, regioregularity, and side chain chemistry can modulate the inter- and intra-chain interactions and, consequently, affect the chain persistence length. Regioregular conjugated P3HT has been observed to exhibit strong interchain $\pi$-$\pi$ interactions from the planar thiophene rings along the chain backbone.[36] While P3(2MP)T in this study is also regioregular, its side chain contains a branch point, which appears to hinder $\pi$-$\pi$ interactions and allow for more torsional motion between backbone units, resulting in greater chain flexibility.

It is instructive to compare our quantitative conformational characteristics with previous literature results. The persistence length of P3HT suspended in d-DCB determined by Segalman and coworkers[27] appears to be considerably smaller than our results presented in Figure 3b. One possible reason for this discrepancy is the model selected for data analysis. Several theoretical scattering functions have been developed to address the conformation of a "*real*" polymer chain by incorporating the excluded volume effect into the Debye model. Among them is the phenomenological model developed by Sharp and Bloomfield,[37] which was used by Segalman and coworkers in their data analysis. However, computer simulations have demonstrated that this analytical model is only valid up to the spatial range of $Q^2 R_G^2 < 2$ and is therefore unable to reliably resolve local structural details in a quantitative manner. Discussions about the limitation of the Sharp-Bloomfield model can be found elsewhere.[33, 38]

The overall conformation of P3HT and P3(2MP)T in solvent can be further envisioned from the ratio of $b/L$ given in Figure 4. It is found to be around 0.4 and 0.1 for P3HT and P3(2MP)T respectively. This stiff nature of a P3HT chain and the flexibility of a P3(2MP)T chain are described by the corresponding schematic pictures.

In conclusion, we studied the conformation of poly(3-alkylthiophenes), a model conducting conjugated polymer system, dispersed in deuterated dicholorbenzene through small angle neutron scattering. By examining the molecular stiffness, we demonstrated the profound impact of the spatial arrangement of atoms presenting in side chain on the backbone conformation. Now the role of topological isomerism has been identified, perhaps the more difficult problems of how it influences the microscopic mechanism of $\pi$ - $\pi$ interactions and conductivity mechanism can be addressed.

## AUTHOR INFORMATION

**Corresponding Author**

*E-mail: chenw@ornl.gov;

## ACKNOWLEDGMENT

This research at SNS and CNMS of Oak Ridge National Laboratory was sponsored by the Scientific User Facilities Division, Office of Basic Energy Sciences, U.S. Department of Energy. W.D.H. acknowledges the support of ORNL HSRE program.